\documentclass[aps,prb,twocolumn,showpacs,superscriptaddress]{revtex4}
\usepackage{eurosym}
\usepackage{graphicx}
\usepackage{amssymb}
\usepackage{amsmath}
\usepackage{color}

\begin{document}

\title{Exact analytic Gorkov-Ginzburg-Landau theory of type-II
superconductivity in the magneto-quantum oscillations limit}
\author{V. Zhuravlev}
\affiliation{Schulich Faculty of Chemistry, Technion-Israel Institute of Technology,
Haifa 32000, Israel}
\affiliation{Physics Department, Ort Braude College, P.O. Box 78, 21982 Karmiel, Israel }
\author{T. Maniv}
\altaffiliation{e-mail:maniv@tx.technion.ac.il}
\affiliation{Schulich Faculty of Chemistry, Technion-Israel Institute of Technology,
Haifa 32000, Israel}
\date{\today }

\begin{abstract}
A new Green's function representation is employed in a microscopic
derivation of a Ginzburg-Landau theory of strongly type superconductivity at
high magnetic fields. An exact analytical, physically transparent expression
for the quartic term in the corresponding order parameter expansion is
presented. The resulting expression reveals singular non-local contributions
to the superconducting (SC) free energy, associated with highly coherent
cyclotron motions of the paired electrons near the Fermi surface, which are
strongly coupled to the vortex lattice. A major part of these contributions
arises from incoherent scattering by the spatially averaged pair-potential,
which is purely harmonic in the de Haas van Alphen frequency. However,
coherent scatterings by the ordered vortex lattice generate, at low
temperatures, large erratically oscillating (i.e. paramagnetic-diamagnetic)
contribution to the SC free energy as a function of the magnetic field.
Vortex lattice disorder, which tends to suppress this oscillatory component,
is found to preserve the singular harmonic part of the SC free energy.
\end{abstract}

\pacs{74.25.Ha, 74.25.Uv}
\maketitle

\section{Introduction}

Revealing the mechanism in which a clean, ordered superconducting (SC)
material at very low temperatures responds to the application of an external
magnetic field is of crucial importance for understanding the phenomenon of
superconductivity at its most fundamental level. Surprisingly, as compared
to high-field superconductivity under \textquotedblleft
non-ideal\textquotedblright\ conditions, i.e. in dirty or disordered
materials at relatively high temperatures, the \textquotedblleft
ideal\textquotedblright\ SC state at low temperatures under high magnetic
fields is currently not well understood even within the conventional BCS
theory \cite{Rasolt-Tesanovic92},\cite{Maniv01}. Experimental detection of
such high magnetic field states is currently restricted mainly to
magneto-quantum oscillations techniques \cite{Janssen98}, which provide
researchers with detailed quantum information of the system investigated,
but are difficult to analyze by the standard Fourier transform techniques
due to the highly restricted range of magnetic fields available below the SC
transition. \ There have been many reports on observation of de Haas van
Alphen (dHvA) or Shubnikov-de Haas (SdH) oscillations in the SC states of
strongly type-II superconductors, showing occasionally additional damping of
the signal in the SC state with respect to the normal state signal (a
partial list of references includes Refs.\cite%
{Janssen98,Corcoran94,Terashima97,Maniv06,Isshiki08,Bergk09}). \ However,
their interpretation has not lead so far to any kind of consensus regarding
the influence of the vortex matter on the oscillatory signal, partly because
of the difficulties in the data analysis, and partly due to lack of a
consistent theory with a quantitatively predictive power \cite{Maniv01}. The
mean-field theories based on a detailed exposition of the quasi-particle
excitations obtained by solving the corresponding Bogoliubov--de Gennes
(BdG) equations for an ordered vortex lattice \cite%
{Dukan94,Dukan95,Norman95,Norman96,Kita02}, provide insight into fine
features of the Landau band structure, but lose their transparency very
quickly and become heavily numerical at early stages of their application to
any observable quantity. On the other hand, a simple formula for the
additional damping \cite{Maki91,Stephen92}, used frequently in the
literature for a manageable interpretation of experimental data, has been
shown to be limited to situations of random vortex distributions \cite%
{Maniv01}, and to the influence of the SC order parameter on the quasi
particle relaxation rate \cite{Stephen92}, neglecting important
contributions to the oscillatory SC free energy.

Attempting to compare the results of the different theoretical approaches
leads to great confusion. In the field range near $H_{c2}$ where the SC
order parameter is small and the leading (quadratic) term in the order
parameter expansion of the SC free energy should be a good approximation\cite%
{Maniv01}, the results of all numerical simulation of the BdG equations \cite%
{Dukan94,Dukan95,Norman95,Norman96,Kita02} deviate markedly from this
universal limiting value (see e.g. Fig.\ 8 in Ref.\ \cite{Kita02}).The
situation in the low magnetic fields region well bellow $H_{c2}$ is even
worse. \ Norman and MacDonald (NM) in their numerical simulations of the BdG
equations reported \cite{Norman96} that the harmonic of the Fourier
transform of the calculated magnetization in a finite interval of this
regime varies erratically in sign and magnitude (see Fig.5 there) with no
pattern that they could discern. \ Similar calculations carried out by Yasui
and Kita \cite{Kita02} have resulted in an additional damping rate which
deviates drastically from both NM result and Maki's formula, showing
erratically oscillating patterns of the corresponding Dingle plot, which
seems to be indicative of some fundamental problem of numerical instability.

It is therefore desirable to derive an exact analytical expression for the
SC thermodynamic potential which will enables one to identify the origin of
this erratic behavior and carry out the calculation in a controlled fashion.
In this paper we present such an exact analytical expression within the
framework of the Gorkov-Ginzburg-Landau expansion of the free energy in the
SC order parameter (or pair potential) up to fourth order. It is indeed
found that highly correlated portions of the electronic cyclotron orbits
involved in pairing have dominant contributions to the quartic and higher
order terms of the SC free energy. The corresponding, singularly divergent
distribution of the free energy obtained within an effective temporal
representation, results in equally shared contributions from the spatially
uniform component of the SC pair-potential and from its periodically
modulated component associated with the vortex lattice.

The resulting expression of the free energy consists of two types: terms
harmonic in the dHvA frequency and associated with the Landau level
structure of the quasi particles, and erratically oscillating terms as
functions of the magnetic field, associated with coherent scattering
channels of the quasi particles by the vortex lattice. The latter are
strongly enhanced when the length of a reciprocal vortex lattice vector
coincides, or nearly coincides with the Fermi sphere diameter.

The resulting analytical expression also enables one to study the effect of
disorder in the vortex lattice on the oscillatory free energy. In the white
noise limit of the self-consistent Born approximation (SCBA) \cite{Stephen92}%
, the erratic oscillations associated with the coherent scattering by the
vortex lattice are suppressed, yielding a result consistent with Maki's
formula \cite{Maki91}. However, deviations from the SCBA should be carefully
examined in light of the recent high-field low-temperature $\mu $SR
measurements in the vortex-glass (peak-effect) region of borocarbide
superconductors \cite{Maniv11}, which have shown strong correlation betwen
the enhanced additional damping of dHvA oscillations observed in the
peak-effect region with enhanced vortex lattice disorder in this region \cite%
{Maniv01}.

\section{General formulation}

We consider a 2D strongly type-II (neglecting the effect of SC screening
currents) superconductor in a perpendicular uniform magnetic field $\mathbf{%
H=}\left( 0,0,H\right) $. Generalization to isotropic 3D systems is rather
straightforward. It is assumed that the superconductor can be described by
means of BCS-hamiltonian density for the usual singlet $s$-wave electron
pairing

\begin{eqnarray}
\mathcal{H}_{BCS}&=&\Delta ^{\ast }\left( \mathbf{r}\right) \psi \left( 
\mathbf{r}\right) \psi \left( \mathbf{r}\right) +\Delta \left( \mathbf{r}%
\right) \psi ^{\ast }\left( \mathbf{r}\right) \psi ^{\ast }\left( \mathbf{r}%
\right)  \notag \\
&-&\frac{1}{g_{BCS}}\left\vert \Delta \left( \mathbf{r}\right) \right\vert
^{2}
\end{eqnarray}%
where $\psi \left( \mathbf{r}\right) $ is the single electron field operator
and $g_{BCS}$ is BCS coupling constant (electron spin is neglected for the
sake of simplicity). Within mean-field approximation the order parameter, $%
\Delta \left( \mathbf{r}\right) =g_{BCS}\left\langle \psi \left( \mathbf{r}%
\right) \psi \left( \mathbf{r}\right) \right\rangle $, should be determined
self-consistently by minimizing the superconducting thermodynamic potential, 
$\Omega _{sc}\left( \Delta \right) $. We do not consider the problem in a
fully self-consistent manner, assuming that the order parameter is described
by a general vortex lattice state, 
\begin{equation*}
\Delta (\mathbf{r})=\left( \frac{2\pi }{a_{x}^{2}}\right) ^{1/4}\Delta
_{0}\varphi _{0}(\mathbf{r});
\end{equation*}%
written in terms of a discrete set of ground-state Landau orbitals: 
\begin{eqnarray}
\varphi _{0}(x,y) &=&e^{ixy}\sum_{n}e^{-i\theta
n^{2}+iq_{n}x-(y+q_{n}/2)^{2}}  \label{phi0} \\
&=&e^{-\frac{1}{2}|z|^{2}+\frac{1}{2}z^{2}}\sum_{n}e^{iq_{n}z-\frac{q_{n}^{2}%
}{4}}  \notag
\end{eqnarray}%
where $z=x+iy$ and $q_{n}=\frac{2\pi }{a_{x}}n=q_{0}n$ with the lattice
spacing $a_{x}$ along the $x$ - axis and the angular parameter $\theta $%
\textit{\ }which determines the point symmetry of the vortex lattice\textit{%
. }It is easy to see that for a general (rhombic) vortex-lattice geometry,
determined by the angular parameter\textit{\ }$\theta $\textit{\ , }$%
a_{x}^{2}=\pi /\sqrt{1-\left( \theta /\pi \right) ^{2}}$\textit{. }For the
Abrikosov triangular lattice:\textit{\ }$\theta =\pi /2$\textit{\ }and%
\textit{\ }$q_{0}=\frac{2\pi }{a_{x}}=3^{1/4}\sqrt{2\pi }$\textit{.}

We use the symmetric gauge with vector potential $\mathbf{A=}\frac{1}{2}%
\left[ \mathbf{H}\times \mathbf{r}\right] =\frac{1}{2}H\left( -y,x,0\right) $
and dimensionless space coordinates measured in units of the electronic
magnetic length, $a_{H}=\sqrt{c\hbar /eH}$. The amplitude of the order
parameter, $\Delta _{0}^{2}=S^{-1}\int d^{2}\mathbf{r}_{i}\left\vert \Delta (%
\mathbf{r}_{i})\right\vert ^{2},$ where\textit{\ }$S=\pi N$\textit{\ }is the
area of the 2D superconductor and\textit{\ }$N$\textit{\ }is the number of
vortices, is treated as a variational parameter for minimizing $\Omega
_{sc}\left( \Delta \right) $.

The thermodynamic potential, $\Omega _{sc}\left( \Delta \right) $, can be
written as a Taylor expansion in the SC order parameter \cite%
{Rasolt-Tesanovic92}:%
\begin{eqnarray*}
\Omega _{sc}\left( \Delta _{0}\right)  &=&S\frac{\Delta _{0}^{2}}{\left(
\hbar \omega _{c}\right) ^{2}g_{BCS}}+\sum\limits_{n=1}\frac{\left(
-1\right) ^{n}}{n}\Omega _{2n}\left( \Delta _{0}\right) ,\ \ \ \  \\
\Omega _{2n} &=&\Omega _{2n}^{\left( 0\right) }\int d^{2}\left\{ \mathbf{r}%
\right\} \widetilde{\Gamma }_{2n}(\left\{ \mathbf{r}\right\} )\widetilde{K}%
_{2n}(\left\{ \mathbf{r}\right\} )\ \ ,\ \ \ \  \\
\Omega _{2n}^{\left( 0\right) } &=&\left( \frac{2\pi }{a_{x}^{2}}\right)
^{2n/4}\frac{1}{\left( 2\pi \right) ^{2n}}k_{B}Ta_{H}^{2}\left\vert \frac{%
\Delta _{0}}{\hbar \omega _{c}}\right\vert ^{2n}.
\end{eqnarray*}%
where:%
\begin{eqnarray}
&&\widetilde{K}_{2n}(\left\{ \mathbf{r}\right\} )=\left( 2\pi \hbar
^{2}/m\right) ^{2n}\sum_{\nu }G_{0}^{\ast }(\mathbf{r}_{1},\mathbf{r}%
_{2},\omega _{\nu })\times   \label{e-Kernel2n} \\
&&G_{0}(\mathbf{r}_{2},\mathbf{r}_{3},\omega _{\nu })...G_{0}^{\ast }(%
\mathbf{r}_{2n-1},\mathbf{r}_{2n},\omega _{\nu })G_{0}(\mathbf{r}_{2n},%
\mathbf{r}_{1},\omega _{\nu })  \notag
\end{eqnarray}%
\begin{eqnarray}
&&\widetilde{\Gamma }_{2n}(\left\{ \mathbf{r}\right\} )=g^{\ast }(\mathbf{r}%
_{1},\mathbf{r}_{2})g(\mathbf{r}_{2},\mathbf{r}_{3})...g^{\ast }(\mathbf{r}%
_{2n-1},\mathbf{r}_{2n})\times   \label{Vertex2n} \\
&&g(\mathbf{r}_{2n},\mathbf{r}_{1})\varphi _{0}(\mathbf{r}_{1})\varphi
_{0}^{\ast }(\mathbf{r}_{2})...\varphi _{0}(\mathbf{r}_{2n-1})\varphi
_{0}^{\ast }(\mathbf{r}_{2n})  \notag
\end{eqnarray}

Here we use the normal state single electron Green's function in the uniform
magnetic field, which is given by the well known expression,$G(\mathbf{r_{2}}%
,\mathbf{r_{1}},\omega _{\nu })=g(\mathbf{r_{2}},\mathbf{r_{1}})G_{0}(%
\mathbf{r_{2}},\mathbf{r_{1}},\omega _{\nu })$, where: 
\begin{equation}
G_{0}(\mathbf{r_{2}},\mathbf{r_{1}},\omega _{\nu })\equiv G_{0}\left( 
\mathbf{\rho },\omega _{\nu }\right) =\frac{1}{2\pi a_{H}^{2}}\sum_{n}\dfrac{%
e^{-\rho ^{2}/4}L_{n}(\rho ^{2}/2)}{\mu _{F}-\varepsilon _{n}+i\hbar \omega
_{\nu }}\ \ \ \ \ \   \label{GF1}
\end{equation}%
$\mathbf{\rho =r_{2}}-\mathbf{r_{1}}$ , and $g(\mathbf{r_{2}},\mathbf{r_{1}}%
)=e^{-\frac{i}{2}\left[ \mathbf{r_{2}}\times \mathbf{r_{1}}\right] }$ is the
usual gauge factor. Also note that in the above equations $\omega _{\nu
}=\pi k_{B}T\left( 2\nu +1\right) /\hbar ,\nu =0,\pm 1,...$ is the Matsubara
frequency at temperature $T$, $\mu _{F}$ is the chemical potential, and $%
\varepsilon _{n}=\hbar \omega _{c}\left( n+1/2\right) ,$ $n=0,1,2,...$is a
Landau level energy with $\omega _{c}=eH/mc$ the cyclotron frequency.

It will be very helpful to use an equivalent representation of the
translational invariant part of the Green's function for $\omega _{\nu }>0$%
,i.e.: 
\begin{eqnarray}
G_{0}\left( \rho \right) &=&\frac{1}{2\pi \hbar \omega _{c}}\int_{0}^{\infty
}d\tau e^{i\tau \left[ n_{F}+i\varpi _{\nu }\right] }\frac{\exp \left( -%
\frac{\rho ^{2}}{4}\frac{1+e^{-i\tau }}{1-e^{-i\tau }}\right)}{1-e^{-i\tau }}
\notag \\
&=&\frac{1}{2\pi \hbar \omega _{c}}\int_{0}^{\infty }\frac{d\tau }{\alpha }%
e^{i\tau \left[ n_{F}+i\varpi _{\nu }\right] -\mu \rho ^{2}}
\label{G0thauRep}
\end{eqnarray}%
which can be easily derived from Eq.\ref{GF1} by using the integral
representation of $\left( \mu _{F}-\varepsilon _{n}+i\hbar \omega _{\nu
}\right) ^{-1}$ and the generating function of the Laguerre polynomials. The
resulting expression is written in terms of the following dimensionless
quantities: $\alpha \equiv 1-e^{-i\tau },\ \ \mu \equiv \frac{1}{4}\frac{%
1+e^{-i\tau }}{1-e^{-i\tau }},$ $\left( \mu _{F}-\varepsilon _{n}\right)
=\hbar \omega _{c}\left( n_{F}-n\right) $ \ \ and\ \ $\omega _{\nu }=\hbar
\omega _{c}\varpi _{\nu }$. Note also that for $\varpi _{\nu }<0$, $\tau $
in Eq.\ref{G0thauRep} should be replaced with $-\tau $ , yielding the
complex conjugate of the expression for $\varpi _{\nu }>0$.

Exploiting the integral representation, Eq.\ref{G0thauRep},\ we can rewrite
the electronic kernel $\widetilde{K}_{2n}(\left\{ \mathbf{r}\right\} )$ in
the form:

\begin{eqnarray*}
&&\widetilde{K}_{2n}(\left\{ \mathbf{r}\right\}
)=\prod\limits_{j}\int_{0}^{\infty }d\tau _{j}e^{-i\varepsilon _{j}\tau
_{j}n_{F}-\varpi _{\nu }\tau _{j}}\frac{1}{\alpha _{j}}\exp \left( -\mu
_{j}\rho _{j}^{2}\right) , \\
&&\varepsilon _{j}=\left( -1\right) ^{j+1},\text{ \ }\alpha
_{j}=1-e^{i\varepsilon _{j}\tau _{j}},\text{ \ }\mu _{j}=\frac{1}{4}\frac{%
1+e^{i\varepsilon _{j}\tau _{j}}}{1-e^{i\varepsilon _{j}\tau _{j}}},
\end{eqnarray*}

This representation of $\Omega _{sc}\left( \Delta _{0}\right) $ has an
obvious advantage over the original expression: all space integrals are of
Gaussian forms and, therefore, can be calculated analytically.
Unfortunately, gauge factors mix all electron coordinates so that the
calculation of the higher order terms is not trivial.

\section{Quartic versus quadratic terms: Effect of the vortex lattice}

\subsection{The quadratic term}

The second order term have been calculated long ago. We repeat the
calculation to illustrate the advantage of using the Green's function in the
special representation, Eq.\ref{G0thauRep}.

The vertex part in Eq.\ref{Vertex2n} can be written as%
\begin{eqnarray*}
\widetilde{\Gamma }_{2}(\mathbf{r}_{1},\mathbf{r}_{2}) &=&g^{\ast }(\mathbf{r%
}_{1},\mathbf{r}_{2})g(\mathbf{r}_{2},\mathbf{r}_{1})\varphi _{0}(\mathbf{r}%
_{1})\varphi _{0}^{\ast }(\mathbf{r}_{2}) \\
&=&\sum\limits_{n,m=-\infty }^{\infty }e^{\zeta _{nm}^{\left( 2\right) }}
\end{eqnarray*}%
where $\zeta _{nm}^{\left( 2\right) }=i\left( x_{1}y_{2}-y_{1}x_{2}\right)
+ix_{1}y_{1}-ix_{2}y_{2}+iq_{n}x_{1}-(y_{1}+q_{n}/2)^{2}-iq_{m}x_{2}-(y_{2}+q_{m}/2)^{2} 
$. Noting that the dependence on the center of mass coordinates, $\mathbf{%
R\equiv }\frac{\mathbf{r}_{1}+\mathbf{r}_{2}}{2}$,appears only in the vertex
part, one can extract this dependence from $\zeta _{nm}^{\left( 2\right) }$,
ending with two integrals over the center of mass: $\int dR_{x}\exp \left[
i\left( q_{n}R_{x}-q_{m}R_{x}\right) \right] =a_{x}N_{x}\delta _{nm}$ and $%
\int dR_{y}^{\prime }\exp \left( -2R_{y}^{\prime 2}\right) =\sqrt{\frac{\pi 
}{2}}.$ Here $N_{x}$ is the number of vortices along the $x$-direction and $%
R_{y}^{\prime }$ is a shifted $R_{y}$ coordinate. The remaining function, $%
\zeta _{nm}^{\left( 2\right) }\rightarrow -\frac{1}{2}\rho ^{2}$ does not depend on $n$%
, so that summation over $n$ gives the number of Landau orbitals along the $%
y $-axis, $N_{y}$. Consequently, the quadratic term can be written as%
\begin{eqnarray}
\Omega _{2\nu } &=&a_{x}\sqrt{\frac{\pi }{2}}N_{x}N_{y}\Omega _{2}^{\left(
0\right) }\int_{0}^{\infty }\int_{0}^{\infty }d\tau _{1}d\tau _{2}
\label{QT} \\
&&e^{in_{F}\left( \tau _{2}-\tau _{1}\right) -\varpi _{\nu }\left( \tau
_{1}+\tau _{2}\right) }\times  \notag \\
&&\frac{1}{\alpha _{1}\alpha _{2}}\int d^{2}\rho \exp \left( -\mu _{1}\rho
^{2}-\mu _{2}\rho ^{2}-\frac{1}{2}\rho ^{2}\right) \ \ \ \   \notag
\end{eqnarray}%
or, after integrating over relative coordinates, $\mathbf{\rho }$, as%
\begin{eqnarray}
\Omega _{2\nu } &=&a_{x}\sqrt{\frac{\pi }{2}}N\Omega _{2}^{\left( 0\right)
}\int_{0}^{\infty }\int_{0}^{\infty }d\tau _{1}d\tau _{2}  \label{QT-fin} \\
&&e^{in_{F}\left( \tau _{2}-\tau _{1}\right) -\varpi _{\nu }\left( \tau
_{1}+\tau _{2}\right) }\frac{1}{\alpha _{1}+\alpha _{2}}  \notag
\end{eqnarray}%
where $N=N_{x}N_{y}$ is a number of vortices in the system.

The dominant contributions to the $\tau $-integrals originates in the poles
of the integrand where $\alpha _{1,2}\rightarrow 0$, namely at $\tau
_{j}\rightarrow 2n_{j}\pi ,n_{j}=0,\pm 1,..$, where the first exponent $%
in_{F}\left( \tau _{2}-\tau _{1}\right) $ is equal to $2i\pi n_{F}n,n=0,\pm
1,..$, corresponding to exact harmonics of the dHvA frequency $F=n_{F}H$. \
We therefore conclude that the quadratic term is dominated by harmonics of
the dHvA frequency which implies that to leading order in the GGL expansion
the Landau levels structure is not distorted by the vortex lattice. This
result is consistent with the well known property of the quadratic term to
be independent of the vortex lattice structure. \ 

Considering the first harmonic for the sake of illustration, we shift $\tau
_{2}\rightarrow 2\pi +\tau _{2}$ and expand\textit{\ }$\alpha _{1}+\alpha
_{2}$\textit{\ }in $\tau _{1}$\textit{\ }and\textit{\ }$\tau _{2}$\textit{\ }%
for $\tau _{1}\ll 1$\ and $\left\vert \tau _{2}\right\vert \ll 1$: $\alpha
_{1}+\alpha _{2}\simeq i\left( \tau _{2}-\tau _{1}\right) +\frac{1}{4}\left(
\tau _{1}+\tau _{2}\right) ^{2}$. Here the term $\frac{1}{4}\left( \tau
_{1}-\tau _{2}\right) ^{2}$\ was neglected since $\left( \tau _{2}-\tau
_{1}\right) \sim \left( \tau _{1}+\tau _{2}\right) ^{2}$\ . Noting, further,
that if $\tau _{1}+\tau _{2}<0$\ the pole is located out of the integration
interval, and calculating the corresponding Cauchy integral over the $\left(
\tau _{2}-\tau _{1}\right) $-variable for $\tau _{1}+\tau _{2}\geq 0$, one
obtains: $\frac{1}{2}k_{B}Ta_{H}^{2}N\frac{\pi ^{3/2}}{\sqrt{n_{F}}}\left( 
\frac{\Delta _{0}}{\hbar \omega _{c}}\right) ^{2}e^{2i\pi n_{F}-2\pi \varpi
_{\nu }}$. A similar expression can be derived by expanding near the
symmetric point $\tau _{1}\rightarrow 2\pi +\tau _{1}$\ and $\tau
_{2}\rightarrow \tau _{2}$\ with $\left\vert \tau _{1}\right\vert \ll 1$\
and $\tau _{2}$\ $\ll 1$. Therefore, the quadratic term is written as 
\begin{equation*}
\Omega _{2}^{\left( 1h\right) }=k_{B}Ta_{H}^{2}N\frac{\pi ^{3/2}}{\sqrt{n_{F}%
}}\left( \frac{\Delta _{0}}{\hbar \omega _{c}}\right) ^{2}\mathit{{Re}%
e^{2i\pi n_{F}-2\pi \varpi _{\nu }}}
\end{equation*}%
\ Since near the poles $\mu _{j}\sim \frac{1}{\tau _{j}}\gg 1$, the spatial
integral in Eq.\ref{QT} is dominated by very small distances, a result
consistent with the locality of the quadratic term. Also note that the final
expression does not depend on $a_{x}$ , a result consistent with the fact
that the structure of the vortex lattice does not influence the quadratic
term.

\subsection{The quartic term}

\subsubsection{Useful analytical expressions\ }

The calculation of the next order term, the quartic term, is much more
complicated since, unlike the quadratic term, it is strongly affected by the
coupling of the electrons to the vortex lattice. However, the use of the
representation, Eq.\ref{G0thauRep}, for the single electron Green's
functions facilitates greatly the entire 8-fold spatial integration by
transforming the corresponding integrand into a multiple Gaussian form. \
Following the derivation described in detail in Appendix A the quartic term
can be written as a 4D 'temporal' integral:%
\begin{eqnarray}
\Omega _{4\nu } &=&\Omega _{4}^{\left( 0\right) }L_{x}N_{y}\frac{\sqrt{\pi }%
}{2}\left\vert \det M^{-1}\right\vert ^{2}\int_{0}^{\infty }d\tau _{1}d\tau
_{2}d\tau _{3}d\tau _{4}  \label{QT2} \\
&&e^{-\varpi _{\nu }\left( \tau _{1}+\tau _{2}+\tau _{3}+\tau _{4}\right)
-in_{F}\left( \tau _{1}-\tau _{2}+\tau _{3}-\tau _{4}\right) }\Phi \left[
\tau \right] L\left[ \tau \right] \ \ \ \   \notag
\end{eqnarray}%
where

\begin{eqnarray}
\Phi \left[ \tau \right] &=&\frac{1}{\alpha _{1}\alpha _{2}\alpha _{3}\alpha
_{4}}\frac{\pi }{\beta _{0}}\frac{\pi ^{2}}{\det U},\ \   \label{phi(thau)}
\\
L\left[ \tau \right] &=&\sum_{st}\exp \left[ -\frac{1}{4}q_{0}^{2}\left(
s^{2}+t^{2}\right) +\frac{1}{4}L^{T}{U}^{-1}L\right] \ \ \ 
\label{L(thau)}
\end{eqnarray}%
where the vector $L$ is given in Eq.\ref{Z and L}, and the matrix $U$ in Eq.%
\ref{U-matrix}.

The calculation of $\det U$ \ can be done by using the relations $\mu _{i}+%
\frac{1}{4}=\frac{1}{2\alpha _{i}}$, and noting that it can be factorized
to: $\det U=\lambda _{b}\lambda _{a}$ where $\lambda _{a}=\frac{1}{2\beta }%
\left( \lambda _{3}-\lambda _{2}\right) $, \ $\lambda _{b}=\frac{1}{2\beta }%
\left( \lambda _{3}+\lambda _{2}\right) $ with $\beta =\frac{1}{2}\left( 
\frac{1}{\alpha _{1}}+\frac{1}{\alpha _{2}}+\frac{1}{\alpha _{3}}+\frac{1}{%
\alpha _{4}}\right) $, $\lambda _{3}=\frac{1}{\alpha _{1}\alpha _{2}\alpha
_{3}}+\frac{1}{\alpha _{2}\alpha _{3}\alpha _{4}}+\frac{1}{\alpha _{1}\alpha
_{2}\alpha _{4}}+\frac{1}{\alpha _{1}\alpha _{3}\alpha _{4}}$, $\ \lambda
_{2}=\frac{1}{\alpha _{1}\alpha _{3}}-\frac{1}{\alpha _{2}\alpha _{4}}$.
Substituting these values to $\Phi \left[ \tau \right] $ one arrives at the
compact expression:%
\begin{equation}
\Phi \left[ \tau \right] =\frac{2\pi ^{3}}{\alpha _{1}+\alpha _{2}+\alpha
_{3}+\alpha _{4}}\frac{1}{\left( 1-\gamma ^{2}\right) ^{1/2}}\ \ \ \ \ \ 
\label{phi}
\end{equation}%
with%
\begin{equation}
\gamma =\frac{\lambda _{2}}{\lambda _{3}}=\frac{\alpha _{2}\alpha
_{4}-\alpha _{1}\alpha _{3}}{\alpha _{1}+\alpha _{2}+\alpha _{3}+\alpha _{4}}%
.  \label{gamma}
\end{equation}

The calculation of the exponential term results in 
\begin{eqnarray}
\frac{1}{4}L^{T}\overline{U}^{-1}L &=&-q_{0}^{2}\left[ \frac{\lambda
_{a}-\lambda _{b}}{\lambda _{b}}s^{2}+\frac{\lambda _{b}-\lambda _{a}}{%
\lambda _{a}}t^{2}\right]  \notag \\
&=&2q_{0}^{2}\gamma \left( \frac{s^{2}}{\gamma +1}+\frac{t^{2}}{\gamma -1}%
\right) .  \label{L}
\end{eqnarray}

Substituting Eqs. \ref{phi} and \ref{L} to Eq. \ref{QT2} we obtain the final
result 
\begin{eqnarray}
\Omega _{4\nu } &=&\frac{1}{2}k_{B}Ta_{H}^{2}N\left\vert \frac{\Delta _{0}}{%
\hbar \omega _{c}}\right\vert ^{4}I_{4},  \label{QT3} \\
I_{4} &=&\int_{0}^{\infty }d\tau _{1}d\tau _{2}d\tau _{3}d\tau _{4}  \notag
\\
&&e^{-\varpi _{\nu }\left( \tau _{1}+\tau _{2}+\tau _{3}+\tau _{4}\right)
-in_{F}\left( \tau _{1}-\tau _{2}+\tau _{3}-\tau _{4}\right) }\times  \notag
\\
&&\frac{\beta \left( \gamma \right) }{\alpha _{1}+\alpha _{2}+\alpha
_{3}+\alpha _{4}}\ \ \ \   \notag
\end{eqnarray}%
where:%
\begin{eqnarray}
\beta \left( \gamma \right) &=&\frac{\sqrt{\pi }}{a_{x}}\frac{1}{\left(
1-\gamma ^{2}\right) ^{1/2}}\times  \label{beta1} \\
&&\sum_{st}\exp \left[ -2i\theta st-\frac{1}{4}q_{0}^{2}\left( \frac{%
1-\gamma }{1+\gamma }s^{2}+\frac{1+\gamma }{1-\gamma }t^{2}\right) \right] \
\ \ \ \ \ \   \notag
\end{eqnarray}

\subsubsection{Major analytical properties}

The structure function $\beta \left( \gamma \right) $, expressed in Eq.\ref%
{beta1}, controls the coupling between the four electrons involved and the
vortex lattice. Its most remarkable feature is associated with the dual
singular points at $\gamma \rightarrow \pm 1$, where the lattice sums over $%
s $ or $t$ \ (depending on whether $1-\gamma \rightarrow 0$ or $1+\gamma
\rightarrow 0$ , respectively) can be replaced by integrals (over $%
\widetilde{s}=s\sqrt{\left( 1-\gamma \right) /2}$ or $\widetilde{t}=t\sqrt{%
\left( 1+\gamma \right) /2}$, respectively), enhancing the singularities of
the corresponding pre-exponential factors to simple poles:

\begin{eqnarray*}
&&\beta \left( \gamma \rightarrow 1\right) \rightarrow \frac{\sqrt{\pi }}{%
a_{x}}\frac{1}{\left( 1-\gamma \right) }\int d\widetilde{s}e^{-\frac{1}{4}%
q_{0}^{2}\widetilde{s}^{2}}\times \\
&&\sum_{t}\exp \left[ -\frac{1}{2}q_{0}^{2}\left( \frac{1}{1-\gamma }\right)
t^{2}-i\sqrt{2}\theta \widetilde{s}\left( \frac{1}{1-\gamma }\right) ^{1/2}t%
\right] \\
&&\rightarrow \frac{1}{\left( 1-\gamma \right) }, \\
&&\beta \left( \gamma \rightarrow -1\right) \rightarrow \frac{\sqrt{\pi }}{%
a_{x}}\frac{1}{\left( 1+\gamma \right) }\int d\widetilde{t}e^{-\frac{1}{4}%
q_{0}^{2}\widetilde{t}^{2}}\times \\
&&\sum_{s}\exp \left[ -\frac{1}{2}q_{0}^{2}\left( \frac{1}{1+\gamma }\right)
s^{2}-i\sqrt{2}\theta \widetilde{t}\left( \frac{1}{1+\gamma }\right) ^{1/2}s%
\right] \\
&&\rightarrow \frac{1}{\left( 1+\gamma \right) }
\end{eqnarray*}

Note that in the sum over the remaining variable, $t$ or $s$, only the
single term $t\left( \text{or }s\right) =0$ survives, due to the large
negative real-part values of the corresponding exponent.\ \ 

It is interesting to note that the values of the individual electronic
"time" variables, $\tau _{j}$, satisfying the singular conditions, $\gamma
\rightarrow \pm 1$, are given, respectively, by: 
\begin{equation}
\tau _{1}=\tau _{3}\rightarrow 0,\tau _{2}\rightarrow n\pi -\tau ,\tau
_{4}\rightarrow n\pi +\tau  \label{gamma=1}
\end{equation}
or \ 
\begin{equation}
\tau _{1}\rightarrow n\pi -\tau ,\tau _{3}\rightarrow n\pi +\tau ,\tau
_{2}=\tau _{4}\rightarrow 0  \label{gamma=-1}
\end{equation}
where $\tau $ is an arbitrary real number in the interval: $-\pi \leq \tau
\leq \pi $, and $n=0,\pm 1,\pm 2,...$. Thus, the electrons at such highly
correlated pairs of cyclotron orbits are resonantly coupled to the entire
vortex lattice, yielding only purely harmonic contributions to the SC free
energy in the dHvA frequency $F=n_{F}H$ since under these conditions: $%
e^{-in_{F}\left( \tau _{1}-\tau _{2}+\tau _{3}-\tau _{4}\right) }\rightarrow
e^{-2\pi inn_{F}}$. \ Note also that at the singular points the factor $%
e^{-\varpi _{\nu }\left( \tau _{1}+\tau _{2}+\tau _{3}+\tau _{4}\right) }$
is equal to $e^{-2\pi \left\vert n\varpi _{\nu }\right\vert }$, determining
the thermal damping of the quantum oscillatory part of the SC free energy,
and a natural (thermal) cutoff for the integrals over $\tau _{j}$.

Another type of singularities of the integrand in Eq.\ref{QT3} corresponds
to the vanishing denominator $\alpha _{1}+\alpha _{2}+\alpha _{3}+\alpha
_{4} $ , which takes place at simultaneous vanishing of all $\alpha
_{j}=1-e^{i\varepsilon _{j}\tau _{j}}$ , namely when $\tau _{j}\rightarrow
2\pi n_{j},n_{j}=0,1,2,...$ \ At the corresponding poles the effective
coupling parameter of the electrons to the vortex lattice $\gamma
\rightarrow 0$ , and one recovers the well known local approximation in
which the electrons are only weakly coupled to the vortex lattice.

\subsubsection{Effect of the vortex lattice}

The effect of the vortex lattice on the free energy can be expressed more
clearly by transforming the lattice double sum in Eq.\ref{beta1} into a 2D
reciprocal vortex lattice summation. \ To do so the summation over $t$ is
transformed by means of Poisson formula into: 
\begin{eqnarray}
&&\sum_{m=-\infty }^{\infty }\int_{-\infty }^{\infty }dt\exp \left[ 2i\left(
\pi m-\theta s\right) t-\left( \frac{\pi }{a_{x}}\right) ^{2}\left( \frac{%
1+\gamma }{1-\gamma }\right) t^{2}\right]  \notag \\
&&=\sqrt{\frac{a_{x}^{2}}{\pi }}\left( \frac{1-\gamma }{1+\gamma }\right)
^{1/2} \times  \notag \\
&&\sum_{m=-\infty }^{\infty }\exp \left\{ -\left( \frac{1-\gamma }{1+\gamma }%
\right) \left( \pi m-\theta s\right) ^{2}\left( \frac{a_{x}}{\pi }\right)
^{2}\right\}  \notag
\end{eqnarray}
, so that: 
\begin{eqnarray}
&&\beta \left( \gamma \right) = \frac{1}{\left( 1+\gamma \right) }%
\sum_{s,m=-\infty }^{\infty }  \label{betagamma} \\
&&\exp \left\{ -\left( \frac{1-\gamma }{1+\gamma }\right) \left[ \left( 
\frac{\pi }{a_{x}}\right) ^{2}s^{2}+\left( \pi m-\theta s\right) ^{2}\left( 
\frac{a_{x}}{\pi }\right) ^{2}\right] \right\}  \notag
\end{eqnarray}

Now, using two primitive vectors spanning the vortex lattice: $\mathbf{a}=%
\widehat{x}a_{x}$ \ , $\ \mathbf{b}=\widehat{x}b_{x}+\widehat{y}b_{y}$ ,
with\ $b_{y}=\pi /a_{x}$ , the corresponding primitive vectors spanning the
reciprocal vortex lattice are: $\mathbf{a}^{\ast }=\widehat{x}b_{y}-\widehat{%
y}b_{x},\mathbf{b}^{\ast }=\widehat{y}a_{x}$, so that 
\begin{equation*}
\left( \frac{\pi }{a_{x}}\right) ^{2}s^{2}+\left( \theta s-\pi m\right)
^{2}\left( \frac{a_{x}}{\pi }\right) ^{2}=\left( s\mathbf{a}^{\ast }+m%
\mathbf{b}^{\ast }\right) ^{2}
\end{equation*}%
and 
\begin{eqnarray}
\beta \left( \gamma \right) &=&\frac{1}{\left( 1+\gamma \right) }%
\sum_{s,m=-\infty }^{\infty }\exp \left[ -\left( \frac{1-\gamma }{1+\gamma }%
\right) \left\vert \mathbf{G}_{sm}\right\vert ^{2}\right] ,  \notag \\
\mathbf{G}_{sm} &\equiv &s\mathbf{a}^{\ast }+m\mathbf{b}^{\ast }
\label{beta-Recip}
\end{eqnarray}

A similar procedure in which Poisson formula is used with respect to the
summation over $s$ leads to an expression identical to Eq.\ref{beta-Recip}
after exchanging $\gamma \longleftrightarrow -\gamma $. \ Since under the
integrations over all $\tau _{j}$\ , $j=1,...,4$ \ , a given value of\ $%
\gamma $ (corresponding to a given selection of $\tau _{1},\tau _{3},\tau
_{2},\tau _{4}$ ) always appears with its opposite value $-\gamma $
(corresponding to the values of $\tau _{1},\tau _{3}$ exchanged with those
of\ $\tau _{2},\tau _{4}$) one can always replace $\beta \left( \gamma
\right) $ in Eq.\ref{QT3} with the symmetric expression:

\begin{eqnarray}
&&\beta ^{sym}\left( \gamma \right) = \sum_{\mathbf{G}}\beta _{\mathbf{G}%
}^{sym}\left( \gamma \right) =  \label{beta-RecipSym} \\
&&\frac{1}{2}\sum_{\mathbf{G}}\left\{ \frac{1}{\left( 1+\gamma \right) }\exp %
\left[ -\left( \frac{1-\gamma }{1+\gamma }\right) \left\vert \mathbf{G}%
\right\vert ^{2}\right] \right.  \notag \\
&&\left. +\frac{1}{\left( 1-\gamma \right) }\exp \left[ -\left( \frac{%
1+\gamma }{1-\gamma }\right) \left\vert \mathbf{G}\right\vert ^{2}\right]%
\right\}  \notag
\end{eqnarray}%
without altering the result of $I_{4}$. \ 

Near the singular points $\gamma =\pm 1$ Eq.\ref{beta-RecipSym} describes
two additive coherent processes of two electron pairs moving in cyclotron
orbits on the Fermi surface and undergoing scatterings by the vortex
lattice. Near the singular point $\gamma \rightarrow 1$, where the positions
of the electrons labeled $\left( 1,3\right) $ along their cyclotron orbit
coincide (i.e. for $\tau _{1}=\tau _{3}\rightarrow 0$, see Eq.\ref{gamma=1}%
), the other two electrons, labeled $\left( 2,4\right) $, are moving
coherently along their cyclotron orbits in opposite directions (i.e. $\tau
_{2}\rightarrow n\pi -\tau ,\tau _{4}\rightarrow n\pi +\tau $, see Eq.\ref%
{gamma=-1}). Thus, the singular $\gamma \rightarrow 1$ contribution of the
first term within the brackets in Eqs.\ref{beta-RecipSym} is associated with
the electrons labeled $\left( 1,3\right) $ and involves many $G$-vectors,
whereas the singular $\gamma \rightarrow 1$ contribution of the second term
is associated with the other two electrons labeled $\left( 2,4\right) $ and
involves only the $G=0$ channel. Similarly,\ near the dual singular point $%
\gamma \rightarrow -1$, where the $\left( 1,3\right) $ electrons are moving
in opposite directions (i.e. $\tau _{1}\rightarrow n\pi -\tau $, $\tau
_{3}\rightarrow n\pi +\tau $ ) and the positions of the $\left( 2,4\right) $
electrons along their orbit coincide (i.e. $\tau _{2}=\tau _{4}\rightarrow 0$
), the contribution of the first term involves only the $G=0$ channel,
whereas the contribution of the second term involves many $G$-vectors.

The physical meaning of the singular $\gamma \rightarrow \pm 1$
contributions is therefore apparent: The two electrons whose positions on
the cyclotron orbit coincide at the singular point undergo local mutual
scattering and so exchange many $G$-vectors through the vortex lattice
during the scattering process, while those electrons moving coherently on a
large cyclotron orbit in opposite directions are mutually scattered through
the entire vortex lattice, and so do not exchange momentum.

The resulting leading contributions to\emph{\ }$\beta ^{sym}\left( \gamma
\right) $\emph{\ }can be therefore written in terms of very simple formulas:
The\ forward scattering contribution takes the form: 
\begin{equation}
\beta _{\mathbf{G=0}}^{sym}\left( \gamma \right) =\frac{1}{2}\left( \frac{1}{%
1+\gamma }+\frac{1}{1-\gamma }\right)  \label{beta_g=0}
\end{equation}%
whereas the rest of the reciprocal lattice contributions,$\sum_{\mathbf{%
G\neq 0}}\beta _{\mathbf{G}}^{sym}\left( \gamma \right) $, which involve
increasingly large numbers of reciprocal lattice vectors as $\gamma
\rightarrow \pm 1$, can be well approximated in these limiting cases by the
two-dimensional integral: $\int \beta _{\mathbf{G}}^{sym}\left( \gamma
\right) d^{2}G$, yielding:

\begin{equation}
\sum_{\mathbf{G\neq 0}}\beta _{\mathbf{G}}^{sym}\left( \gamma \right)
\rightarrow \int \beta _{\mathbf{G}}^{sym}\left( \gamma \right) d^{2}G=\frac{%
1}{2}\left\{ 
\begin{array}{c}
\frac{1}{1-\gamma }\text{ \ , \ }\gamma \rightarrow 1 \\ 
\frac{1}{1+\gamma }\text{ \ , \ }\gamma \rightarrow -1%
\end{array}%
\right\}  \label{beta-nonzerogs}
\end{equation}

Note that the $G=0$ term, given by Eq. \ref{beta_g=0}, represents the effect
of the spatially uniform component of the SC order parameter on the free
energy whereas the rest of the terms in Eq. \ref{beta-RecipSym} correspond
to all possible umklapp (coherent) scattering processes by the vortex
lattice.

To gain further insight into this remarkable coupling to the vortex lattice
we may expand $\beta \left( \gamma \right) /\left( \alpha _{1}+\alpha
_{2}+\alpha _{3}+\alpha _{4}\right) $ about one of the singular points, say $%
\gamma =1$, and carry out the $\tau _{j}$-integrations to derive a more
transparent (but approximate) expression for $I_{4}$. \ Focusing, for
simplicity, on the first harmonic of the dHvA frequency $F=n_{F}H$, our
small expansion parameters are (see also Sec.B4): $\widetilde{\xi }_{1}=%
\frac{1}{4}\left( \tau _{1}+\tau _{2}+\tau _{3}+\tau _{4}\right) -\pi /2,%
\widetilde{\xi }_{2}=\frac{1}{2}\left( \tau _{1}-\tau _{2}+\tau _{3}-\tau
_{4}\right) +\pi ,\widetilde{\xi }_{3}=\tau _{1}-\tau _{3},$ and $\widetilde{%
\xi }_{4}=\tau _{4}-\tau _{2}$ , so that to second order, the key composite
variables are given by: 
\begin{equation}
\frac{1-\gamma }{1+\gamma }\simeq -\frac{1}{4}i\widetilde{\xi }_{2}+\frac{1}{%
16}\left( 4\widetilde{\xi }_{1}^{2}+\widetilde{\xi }_{3}^{2}\right)
\label{expand}
\end{equation}%
and:

\begin{eqnarray*}
I_{4} &\rightarrow &e^{2\pi in_{F}}e^{-2\pi \varpi _{\nu }}\int_{0}^{\infty
}d\widetilde{\xi }_{1}e^{-4\varpi _{\nu }\widetilde{\xi }_{1}}\times \\
&&\sum_{\mathbf{G}}\int_{-2\widetilde{\xi }_{1}}^{2\widetilde{\xi }_{1}}d%
\widetilde{\xi }_{2}\exp \left\{ i\widetilde{\xi }_{2}\left[ \frac{1}{4}%
\left\vert \mathbf{G}\right\vert ^{2}-2n_{F}\right] \right\} \times \\
&&\int_{-\left( 2\widetilde{\xi }_{1}+\widetilde{\xi }_{2}\right) }^{2%
\widetilde{\xi }_{1}+\widetilde{\xi }_{2}}d\widetilde{\xi }_{3}\exp \left\{
-\left( \widetilde{\xi }_{1}^{2}+\frac{1}{4}\widetilde{\xi }_{3}^{2}\right) 
\frac{1}{4}\left\vert \mathbf{G}\right\vert ^{2}\right\} \times \\
&&\int_{-\left( 2\widetilde{\xi }_{1}-\widetilde{\xi }_{2}\right) }^{2%
\widetilde{\xi }_{1}-\widetilde{\xi }_{2}}d\widetilde{\xi }_{4}
\end{eqnarray*}

Considering the umklapp scattering terms with large vectors $\mathbf{G}$ \
it is clear that the dominant contributions originate from reciprocal
lattice vectors satisfying: $\frac{1}{2}\left\vert \mathbf{G}\right\vert
\approx \sqrt{2n_{F}}$, namely having length close to the Fermi surface
diameter. Furthermore, due to the large values of $n_{F}$ and the discrete
nature of $\mathbf{G}$ (which are measured in units of the magnetic length)
with an elementary unit of about $\pi $ , the integration over $\xi _{2}$
yields erratically oscillating function of $n_{F}$, which reflects dramatic
influence of the vortex lattice on the fermionic quasi-particles at high
magnetic field.

\subsubsection{Numerical calculations}

For numerical calculations we use Eq.\ref{QT3} assuming a square vortex
lattice with $a_{x}=\sqrt{\pi }$. Performing Poisson summation over $s$ or $t
$ in Eq.\ref{beta1} one can transform $\beta \left( \gamma \right) $ into
simpler, equivalent forms:%
\begin{eqnarray}
\beta _{sq}\left( \gamma \right)  &=&\frac{1}{1-\gamma }\sum_{mt}\exp \left[
-\pi \frac{1+\gamma }{1-\gamma }\left( m^{2}+t^{2}\right) \right] 
\label{beta_sq_num} \\
&=&\frac{1}{1-\gamma }\left( \sum_{n}\exp \left[ -\pi \frac{1+\gamma }{%
1-\gamma }n^{2}\right] \right) ^{2}  \notag \\
&=&\frac{1}{1+\gamma }\left( \sum_{n}\exp \left[ -\pi \frac{1-\gamma }{%
1+\gamma }n^{2}\right] \right) ^{2}  \notag
\end{eqnarray}

The integrals in Eq.\ref{QT3} can be more conveniently evaluated by
transforming to the new variables (shifted with respect to $\widetilde{\xi }%
_{i}$, defined above Eq.\ref{expand}): $\xi _{1}=\frac{1}{4}\left( \tau
_{1}+\tau _{2}+\tau _{3}+\tau _{4}\right) $, $\xi _{2}=\frac{1}{2}\left(
\tau _{1}-\tau _{2}+\tau _{3}-\tau _{4}\right) $, $\xi _{3}=\left( \tau
_{1}-\tau _{3}\right) $, $\xi _{4}=\left( -\tau _{2}+\tau _{4}\right) $:

\begin{eqnarray}
I_{4} &=&\int_{0}^{\infty }d\xi _{1}e^{-4\varpi _{\nu }\xi _{1}}I_{3}\left(
\xi _{1}\right) ,  \label{I4} \\
I_{3}\left( \xi _{1}\right) &\equiv &\int_{-2\xi _{1}}^{2\xi _{1}}d\xi
_{2}e^{-2in_{F}\xi _{2}}\int_{-\left( 2\xi _{1}+\xi _{2}\right) }^{2\xi
_{1}+\xi _{2}}\times  \label{I3} \\
&&d\xi _{3}\int_{-\left( 2\xi _{1}-\xi _{2}\right) }^{2\xi _{1}-\xi
_{2}}d\xi _{4}\frac{\beta \left( \gamma \right) }{\alpha _{1}+\alpha
_{2}+\alpha _{3}+\alpha _{4}}  \notag
\end{eqnarray}

The distribution function $I_{3}\left( \xi _{1}\right) $, has been
calculated numerically for different integer values of $n_{F}$. \ Selecting
integer values of $n_{F}$ pins the SC free energy at maxima of its magnetic
quantum oscillations, allowing to determine their amplitude for any given
harmonic in the dHvA frequency $F=Hn_{F}$. \ The result for $I_{3}\left( \xi
_{1}\right) $ is shown in Fig.1. It appears as a series of sharp peaks
located around the points $\xi _{1}^{k}=\frac{\pi }{2}k$ with $k=0,1,..,$
having monotonically increasing intensity with increasing order $k$. The
maximum positions of the peaks are slightly shifted with respect to $\frac{%
\pi }{2}k$ toward larger values due to the $\xi _{1}$-dependence of the $\xi
_{3},\xi _{4}$-integrals. The peaks' height is found to increase with
increasing harmonic order $k$ as $k^{2}$, but the number of significantly
contributing peaks is limited by the thermal damping factor $e^{-4\varpi
_{\nu }\xi _{1}}$. \ A simple estimation shows that for $2\pi
^{2}k_{B}T\gtrsim 4\hbar \omega _{c}$ the contribution of the second
harmonic does not exceed 10\% of the first harmonic where the 3rd harmonic
contribution is less than 1\%. \ On the other hand at temperatures as low as 
$2\pi ^{2}k_{B}T\lesssim \hbar \omega _{c}$ many harmonics provide
comparable contributions. In this low temperature limit, replacing summation
over harmonics with integration one finds for an integer $n_{F}$: $%
k_{B}T\sum_{\nu }\int_{0}^{\infty }d\xi _{1}e^{-4\varpi _{\nu }\xi
_{1}}I_{3}\left( \xi _{1}\right) \rightarrow k_{B}T\sum_{\nu ,k}k^{2}\exp %
\left[ -\frac{\pi ^{2}k_{B}T}{\hbar \omega _{c}}\left( 2\nu +1\right) k%
\right] $ $\rightarrow \hbar \omega _{c}\left( \frac{\hbar \omega _{c}}{%
k_{B}T}\right) ^{2}$. Therefore, the quartic term diverges as $\frac{1}{T^{2}%
}$ as $T\rightarrow 0$, due to the resonance pairing conditions \cite%
{Maniv01} characterizing the zero spin splitting situation considered here.
Note, however, that the resulting divergence is weaker than that obtained in
the local approximation \cite{Maniv01}. \ In the latter the quartic term for
an integer $n_{F}$ was found to be proportional to $k_{B}T\sum_{\nu }q_{\nu
}^{2}$, with $q_{\nu }=\frac{e^{X_{\nu }}}{\cosh X_{\nu }+\cos 2\pi \left(
n_{F}+1/2\right) }$, and $\ X_{\nu }=\frac{2\pi ^{2}k_{B}T}{\hbar \omega _{c}%
}\left( 2\nu +1\right) $, with the following low temperature limit: $%
k_{B}T\sum_{\nu }q_{\nu }^{2}\rightarrow \frac{1}{T^{3}}$. \ Note also that
the quadratic term, which is local in nature, is characterized by the low
temperature limit: $k_{B}T\sum_{\nu }q_{\nu }\rightarrow \frac{1}{T}$.

\begin{figure}[tbp]
\includegraphics[scale=.6]{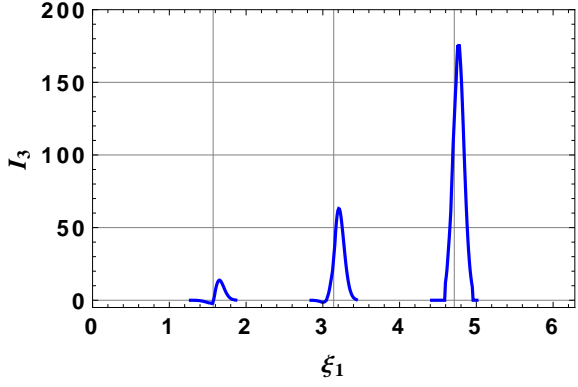}
\caption{The distribution function $I_{3}\left( \protect\xi _{1}\right) $
given by Eq.\protect\ref{I3} for $n_{F}=32$. The peak near $\protect\xi %
_{1}=0$ is too small to be observable in the considered scale .}
\label{fig1}
\end{figure}

For the sake of illustrating the novel (quantum) features of the theory we
will focus here on the leading magnetic quantum oscillatory effect by
considering the first harmonic of the thermodynamic potential in the dHvA
frequency $F=Hn_{F}$. \ This situation corresponds to the usual dHvA
experimental conditions when higher harmonics are relatively small. Under
these circumstances the main contribution to $I_{4}$ (see Eq.\ref{I4})
originates in the second peak at $\xi _{1}\simeq \pi /2$, which is dominated
by the integral over small intervals around $\xi _{2}\simeq \pm \pi $, and
to lesser extent by all other values of $\xi _{2}$. The resulting integral
over $\xi _{2}$ (with the integrand including $e^{-2in_{F}\xi _{2}}$, see Eq.%
\ref{I3}) in the small intervals near $\xi _{2}\simeq \pm \pi $ yield the
dominant contribution to the first harmonic. As usual for the first harmonic
one may restrict the thermal Matsubara summation to the single term $\nu =0$.

The integration over $\xi _{1}$ around the point $\xi _{1}\simeq \pi /2$ has
been performed for different integer values of $n_{F}$ under the assumption
that $e^{-4\varpi _{\nu }\xi _{1}}\simeq e^{-2\pi \varpi _{\nu =0}}$. The
result presented in Fig.2 (blue line) shows clearly the erratic oscillatory
dependence on $n_{F}$ associated with the coupling to the vortex lattice.

\begin{figure}[tbp]
\includegraphics[scale=.6]{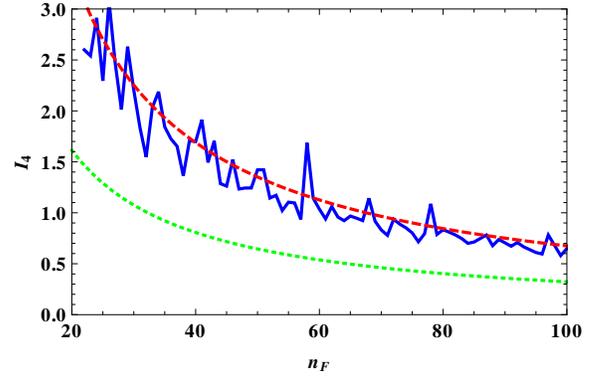}
\caption{The four-fold $\protect\tau $- integral, $I_{4}$, calculated at
integer values of $n_{F}$ for a square vortex lattice (blue solid line). The
red dashed line presents the harmonic part, obtained by using Eq. \protect
\ref{beta_sq_h}. Use of the first term in this expression (corresponding to
the single pole at $\protect\gamma =1$) in the calculation of $I_{4}$ yields
the green dotted line.}
\label{fig2}
\end{figure}

The mean base line of this function, shown in the figure, corresponds to\ $%
I_{4}\left( \xi _{1}\simeq \pi /2\right) $ calculated after replacing $\beta
_{sq}\left( \gamma \right) $ with: 
\begin{equation}
\beta _{sq}^{\left( h\right) }\left( \gamma \right) \equiv \frac{1}{1-\gamma 
}+\frac{1}{1+\gamma },  \label{beta_sq_h}
\end{equation}%
The result is purely harmonic, as can be seen by expanding one of the
denominators, e.g. $\left( \alpha _{1}+\alpha _{2}+\alpha _{3}+\alpha
_{4}\right) \left( 1-\gamma \right) \rightarrow 0$ near $\tau _{1},\tau
_{3}\rightarrow 0;\tau _{2},\tau _{4}\rightarrow \pi $ , in the small
variables ($\widetilde{\tau }_{i}\ll 1$), $\widetilde{\tau }_{1}=\tau _{1},%
\widetilde{\tau }_{2}\rightarrow -\pi +\tau _{2},\widetilde{\tau }%
_{3}\rightarrow \tau _{3},\widetilde{\tau }_{4}\rightarrow -\pi +\tau _{4}$
, up to second order and keeping only leading terms in each variable (see Eq.%
\ref{expand}). The pole contribution at $\xi _{2}$ \ $\left( \xi _{2}=-\pi +%
\widetilde{\xi }_{2}\right) $ yields the first harmonic $e^{2in_{F}\pi
}e^{-2n_{F}\left( \widetilde{\xi }_{1}^{2}+\frac{1}{4}\widetilde{\xi }%
_{3}^{2}\right) }$, which is strongly localized around the origin along both
directions $\xi _{1}$ and $\xi _{3}$ with a characteristic width $\sim \frac{%
1}{\sqrt{n_{F}}}$. The integral over $\xi _{4}$ is not local and it is
restricted only by its integration limits $\pm \left( 2\xi _{1}-\xi
_{2}\right) \simeq \pm 2\pi $. The remaining local (Gaussian) behavior in
the corresponding 2D subspace enables one to estimate the global dependence
of $I_{4}^{\left( 1h\right) }\left( \xi _{1}\simeq \pi /2\right) $ on $n_{F}$
as $I_{4}^{\left( 1h\right) }\left( \xi _{1}\simeq \pi /2\right) \sim \frac{1%
}{n_{F}}$.

Fig.2 also confirms the conclusion drawn in Sec.IIIB3 on the basis of an
analytical consideration saying that umklapp scattering of electron pairs by
the vortex lattice via large reciprocal lattice vectors across the entire
fermi sphere diameter leads to erratic oscillatory dependence of the
thermodynamic potential on $n_{F}=\frac{F}{H}$ about the base line envelope $%
\sim \frac{1}{n_{F}}$. \ The absence of similar Umklapp scattering effects
in the leading, quadratic term in the order parameter expansion, and their
expected increasingly enhanced appearances in higher order terms of this
expansion indicate that the irregularity discussed above should appear
pronounced far from the SC transition where the quartic and higher order
terms become important.

The final result for the first harmonic of the SC thermodynamic potential,
up to fourth order, can be written in the form:%
\begin{eqnarray}
\Omega _{sc}^{\left( 1h\right) }/\Omega _{n}^{\left( 1h\right) } &\simeq &1-%
\frac{\pi ^{3/2}}{\sqrt{n_{F}}}\left\vert \frac{\Delta _{0}}{\hbar \omega
_{c}}\right\vert ^{2}  \label{Omega-Fin} \\
&&+\frac{1}{2}w_{0}\left( 1+w\left( n_{F}\right) \right) \frac{\pi ^{3}}{%
n_{F}}\left\vert \frac{\Delta _{0}}{\hbar \omega _{c}}\right\vert ^{4}-...\
\ \ \ \   \notag
\end{eqnarray}%
where $w_{0}\simeq 1.1$ \ arises from the spatially uniform component of the
SC order parameter, and is purely harmonic, whereas $w\left( n_{F}\right) $,
shown in Fig.3, represents effects of umklapp scattering by the vortex
lattice leading to deviations from the purely harmonic Fourier spectrum.

\begin{figure}[tbp]
\includegraphics[scale=.6]{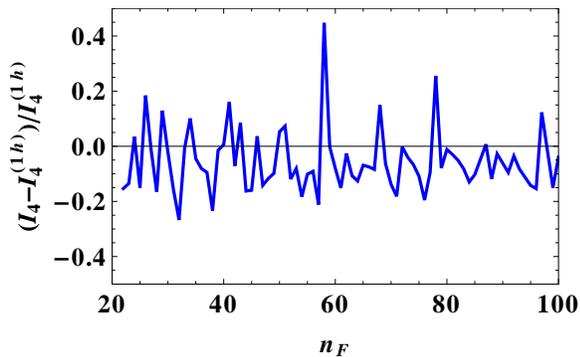}
\caption{The "erratic" function $w\left( n_{F}\right) $ showing the relative
contribution to $I_{4}$ associated with the coherent scattering by the
vortex lattice. \ Note that the negative jumps are due to paramagnetic
distortions of the cyclotron orbits traversing through vortex core regions,
whereas the positive jumps are associated with diamagnetic distortions.}
\label{fig3}
\end{figure}

It is interesting to note that these "erratic" umklapp scattering processes
can be viewed in real space as arising from the passages of paired electrons
in cyclotron orbits (near the fermi energy) through vortex core regions,
where the cyclotron orbit is strongly distorted by the pair-potential into
small circular orbits around the vortex cores \cite{Maniv01}. The resulting
deviations from the normal state cyclotron orbit in a vortex core are
paramagnetic or diamagnetic, depending on the electron energy relative to
the Fermi surface, with the paramagnetic sectors leading to the sharp drops
of the free energy shown in Fig.3, while the diamagnetic ones yielding the
sharp rises seen there.

The existence of these erratic oscillations is due to the highly coherent
cyclotron motions of the two pairs of electrons responsible for the singular
terms $\gamma =\pm 1$ in Eq.\ref{beta-RecipSym}. A scattering process of
these electrons which can destroy this coherence should lead to removal of
the singular behavior. Leaving to future publications the question of how
such scattering processes can be implemented into the present theory (see
the discussion in Sec.V), it is desirable to investigate the robustness of
the quartic term $I_{4}$\ with respect to smearing of the singularities at $%
\gamma =\pm 1$. This can be done by artificially shifting $1-\gamma ,$ and $%
1+\gamma $ in Eq.\ref{beta_sq_num} slightly away form their vanishing forms
to $1+\sigma -\gamma ,$ and $1+\sigma +\gamma $ respectively, for small
values of $\sigma >0$, and repeating the calculation shown in Fig.2. The
result for $\sigma =0.01$ is shown in Fig.4. \ In addition to the
significant reduction of the overall magnitude and suppression of the
(coherent-scattering) "erratic" oscillations, the $n_{F}$ dependence of the
mean base line changes from $n_{F}^{-1}$ to $n_{F}^{-3/2}$, characterizing
the local approximation of the GGL theory \cite{Maniv01}.

\begin{figure}[tbp]
\includegraphics[scale=.6]{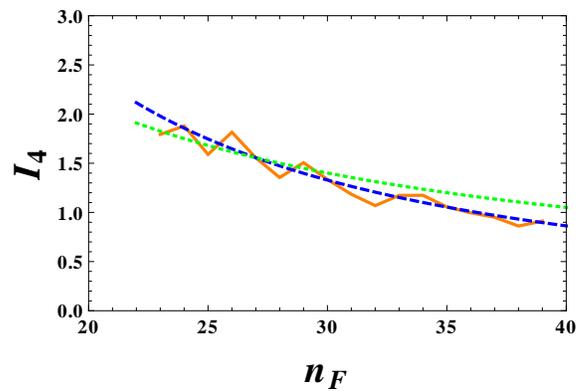}
\caption{$I_{4}$, calculated at integer values of $n_{F}$ for a square
vortex lattice, as in Fig.2, in which the singularities at $\protect\gamma %
=\pm 1$ are removed (see text) with the regularization parameter $\protect%
\sigma =.01$ (red solid line). The (blue) dashed line, which is proportional
to $n_{F}^{-3/2}$, is the best fitting curve, whereas the (green) dotted
curve follows a fitting formula proportional to $n_{F}^{-1}$.}
\label{fig4}
\end{figure}

\section{The effect of vortex lattice disorder in the white noise limit}

Calculation of the influence of vortex-lattice disorder on the SC free
energy in the magneto-quantum oscillations limit can be performed
analytically in the white noise limit. Invoking the general expansion of the
state function $\varphi _{0}(x,y)$ in terms of Landau orbitals wave
functions, $\varphi
_{0}(x,y)=e^{ixy}\sum_{n}c_{n}e^{iq_{n}x-(y+q_{n}/2)^{2}} $, the structure
factor takes the form:

\begin{eqnarray}
\beta \left( \gamma \right) &=&\frac{\sqrt{\pi }}{a_{x}}\frac{1}{\left(
1-\gamma ^{2}\right) ^{1/2}}\frac{1}{N_{x}}\times  \label{beta-Gen} \\
&&\sum_{nst}\exp \left[ -\left( \frac{\pi }{a_{x}}\right) ^{2}\left( \frac{%
1-\gamma }{1+\gamma }s^{2}+\frac{1+\gamma }{1-\gamma }t^{2}\right) \right]
\times  \notag \\
&&c_{n}^{\ast }c_{n+s+t}^{\ast }c_{n+s}c_{n+t}  \notag
\end{eqnarray}%
where the coefficients $\left\{ c_{n}\right\} $ may be considered as random
variables. Averaging the structure factor over realizations of these
coefficients and exploiting the usual (Wick) decoupling: 
\begin{eqnarray*}
&&\left\langle \beta \left( \gamma \right) \right\rangle = \frac{\sqrt{\pi }%
}{a_{x}}\frac{1}{N_{x}\left( 1-\gamma ^{2}\right) ^{1/2}}  \notag \\
&&\sum_{nst}\exp \left[ -\left( \frac{\pi }{a_{x}}\right) ^{2}\left( \frac{%
1-\gamma }{1+\gamma }s^{2}+\frac{1+\gamma }{1-\gamma }t^{2}\right) \right]
\times \\
&&\left[ \left\langle c_{n}^{\ast }c_{n+s}\right\rangle \left\langle
c_{n+s+t}^{\ast }c_{n+t}\right\rangle +\left\langle c_{n+s+t}^{\ast
}c_{n+s}\right\rangle \left\langle c_{n}^{\ast }c_{n+t}\right\rangle \right]
\end{eqnarray*}%
, in the white noise limit, i.e. $\left\langle c_{n}^{\ast
}c_{n+s}\right\rangle \left\langle c_{n+s+t}^{\ast }c_{n+t}\right\rangle
\rightarrow \delta _{n,n+s}\delta _{n+s+t,n+t}=\delta _{s,0}$ , and $\
\left\langle c_{n+s+t}^{\ast }c_{n+s}\right\rangle \left\langle c_{n}^{\ast
}c_{n+t}\right\rangle \rightarrow \delta _{n+s+t,n+s}\delta _{n,n+t}=\delta
_{t,0}$ , one finds:

\begin{equation}
\left\langle \beta \left( \gamma \right) \right\rangle =\frac{\sqrt{\pi }}{%
a_{x}}\frac{1}{\left( 1-\gamma ^{2}\right) ^{1/2}}\left[ 
\begin{array}{c}
\sum_{s}e^{-\left( \frac{\pi }{a_{x}}\right) ^{2}\left( \frac{1-\gamma }{%
1+\gamma }\right) s^{2}} \\ 
+\sum_{t}e^{-\left( \frac{\pi }{a_{x}}\right) ^{2}\left( \frac{1+\gamma }{%
1-\gamma }\right) t^{2}}%
\end{array}%
\right]  \label{beta-decoup}
\end{equation}

The final step in the procedure leading to the white noise limit should be
the replacement of the discrete summations in Eq.\ref{beta-decoup} with
integrations (e.g. by taking $a_{x}\rightarrow \infty $ there), resulting in
the expression:

\begin{equation}
\left\langle \beta \left( \gamma \right) \right\rangle \rightarrow \frac{1}{%
1-\gamma }+\frac{1}{1+\gamma }  \label{ave-beta}
\end{equation}

This is a rather surprising results since it is seen to be twice Eq.\ref%
{beta_g=0}, obtained for the forward scattering term. The latter (i.e. the $%
G=0$ term), which is usually associated with all incoherent scattering
processes, is expected to be the sole survivor of an averaging over
white-noise disorder, and as such to coincide with Eq.\ref{ave-beta}.

In this limiting case, only incoherent scattering processes by the vortex
matter contribute to the SC thermodynamic potential, and the final result,
up to fourth order, is purely harmonic, with the first harmonic given by:%
\begin{equation}
\left\langle \Omega _{sc}^{\left( 1h\right) }\right\rangle /\Omega
_{n}^{\left( 1h\right) }\simeq 1-\frac{\pi ^{3/2}}{\sqrt{n_{F}}}\left\vert 
\frac{\Delta _{0}}{\hbar \omega _{c}}\right\vert ^{2}+\frac{1}{2}w_{0}\frac{%
\pi ^{3}}{n_{F}}\left\vert \frac{\Delta _{0}}{\hbar \omega _{c}}\right\vert
^{4}-...\ \ \ \ \   \label{ave-Omega}
\end{equation}%
i.e., very close to the well known Maki-Stephen expression\cite%
{Maki91,Stephen92}, as expanded to the same order in $\Delta _{0}$.

An interesting question arises here as to wether the white-noise average of
higher order terms in the order-parameter expansion presented in this paper
also agree with the self-consistent Born approximation (SCBA) inherent to
the Maki-Stephen approach \cite{Maniv01}. \ In particular, possible
destruction of the highly coherent motions of the electron pairs responsible
for the singular contributions to the quartic term $I_{4}$ by an infinite
subset of diagrams which are topologically distinct from the quartic
diagram, might lead to significant deviations from the SCBA.

\section{Conclusion and discussion}

A novel Green's function representation is exploited in this paper for a
microscopic derivation of the Ginzburg-Landau theory of strongly type
superconductivity at high magnetic fields. An exact analytical expression
for the quartic term in the corresponding order parameter expansion, having
a physically transparent form, is presented. The resulting expression
reveals singular non-local contributions to the SC thermodynamic potential,
associated with highly coherent cyclotron motions of the paired electrons
near the Fermi surface, which are strongly coupled to the vortex lattice.
The dominant contributions to the SC free energy, arise from incoherent
scattering by the spatially averaged pair-potential, which is purely
harmonic in the dHvA frequency. However, coherent scatterings by the ordered
vortex lattice generate, at low temperatures, erratically oscillating (i.e.
paramagnetic-diamagnetic) contribution to the SC free energy as a function
of the magnetic field, associated with sharp distortions of the large
quasi-particle cyclotron orbits on the Fermi surface traversing through
vortex core regions. Vortex lattice disorder, which tends to suppress this
oscillatory component, is found to simplify considerably the calculation
allowing analytical evaluation of higher order terms in the order-parameter
expansion. \ However, it can be shown that the infinite subset of diagrams
constituting the standard, self consistent Born approximation (SCBA) \cite%
{Stephen92}, exploited in the white noise limit of the discorded vortex
system \cite{Maniv01}, have the same type of singular points as that found
in our calculation of the quartic term. It would be therefore very
interesting to search for, and then evaluate subsets of diagrams,
topologically distinct from those appearing in the SCBA, which might, after
resummation, destroy the highly coherent cyclotron motions responsible for
the above singularities. Physically speaking, it is expected that the effect
of impurity-scattering on the paired electrons, as calculated beyond the
relaxation time approximation, could destroy this coherence. Whether or not
the robustness of this type of singularities with respect to scattering of
quasi particles by a disordered vortex matter is destroyed by going beyond
the SCBA is a crucial question in our understanding of the vortex lattice
disorder on the dHvA oscillations in the SC state \cite{Maniv11}. 
\begin{acknowledgements}
This research was supported by the Israel Science Foundation, by Posnansky
Research fund in superconductivity, and by EuroMagNET under the EU contract
No.\ 228043.
\end{acknowledgements}

\appendix
\section{}

Similar to the calculation of the quadratic term, it is convenient to
introduce the following, center of mass and relative coordinates:%
\begin{eqnarray*}
\mathbf{R} &=&\frac{1}{4}\left( \mathbf{r_{1}+r_{2}+r_{3}+r_{4}}\right) \\
\mathbf{Q} &=&\frac{1}{2}\left( \mathbf{r_{1}-r_{2}+r_{3}-r_{4}}\right) =%
\frac{1}{4}\left( \mathbf{\rho }_{1}\mathbf{-\rho }_{2}\mathbf{+\rho }_{3}%
\mathbf{-\rho }_{4}\right) \\
\mathbf{D} &=&\frac{1}{2}\left( \mathbf{r_{1}-\mathbf{r_{2}}-r_{3}+r_{4}}%
\right) =\frac{1}{2}\left( \mathbf{\rho }_{4}\mathbf{-\rho }_{2}\right) \\
\mathbf{P} &=&\frac{1}{2}\left( \mathbf{r_{1}+\mathbf{\mathbf{r_{2}}}%
-r_{3}-r_{4}}\right) =\frac{1}{2}\left( \mathbf{\rho }_{1}\mathbf{-\rho }%
_{3}\right)
\end{eqnarray*}%
where \ $\mathbf{\rho }_{i}=\mathbf{r_{i}-r_{i-1}}$. This transformation can
be written in the matrix form: $\boldsymbol{X}=M\ast \mathbf{r}$, where $%
\boldsymbol{X}\equiv \left\{ \mathbf{R,Q,D,P}\right\} $\ are four 2D vectors%
\emph{\ }and $M$\ is a $4\times 4$\ matrix with $\left\vert \det
M\right\vert =1/2$.

\begin{widetext}
All ingredients of the quartic term, which depend on the electronic spatial
coordinates, i.e.:%
\begin{eqnarray*}
\Omega _{4} &=&\Omega _{4}^{\left( 0\right) }\int d^{2}\left\{ \mathbf{r}%
\right\} \widetilde{\Gamma }_{4}(\left\{ \mathbf{r}\right\} )\widetilde{K}%
_{4}(\left\{ \mathbf{r}\right\} ),\ \  \\
\Omega _{4}^{\left( 0\right) } &=&\frac{2\pi }{a_{x}^{2}}\frac{1}{\left(
2\pi \right) ^{4}}k_{B}Ta_{H}^{2}\left\vert \frac{\Delta _{0}}{\hbar \omega
_{c}}\right\vert ^{4} \\
\widetilde{K}_{4}(\left\{ \mathbf{r}\right\} ) &=&\int_{0}^{\infty }d\tau
_{1}d\tau _{2}d\tau _{3}d\tau _{4}e^{-i\left( \tau _{1}-\tau _{2}+\tau
_{3}-\tau _{4}\right) n_{F}-\varpi _{\nu }\left( \tau _{1}+\tau _{2}+\tau
_{3}+\tau _{4}\right) }\times \\
&&\frac{1}{\alpha _{1}\alpha _{2}\alpha _{3}\alpha _{4}}\exp \left[ -\left(
\mu _{1}\rho _{1}^{2}+\mu _{2}\rho _{2}^{2}+\mu _{3}\rho _{3}^{2}+\mu
_{4}\rho _{4}^{2}\right) \right] \\
\widetilde{\Gamma }_{4}(\left\{ \mathbf{r}\right\} ) &=&g^{\ast }(\mathbf{r}%
_{1},\mathbf{r}_{2})g(\mathbf{r}_{2},\mathbf{r}_{3})g^{\ast }(\mathbf{r}_{3},%
\mathbf{r}_{4})g(\mathbf{r}_{4},\mathbf{r}_{1})\varphi _{0}(\mathbf{r}_{1})\varphi _{0}^{\ast }(\mathbf{r}_{2})\varphi
_{0}(\mathbf{r}_{3})\varphi _{0}^{\ast }(\mathbf{r}_{4})
\end{eqnarray*}%
will be rewritten now in terms of the new coordinates. Let us start with the
gauge factors, $g^{\star }(\mathbf{r_{1}},\mathbf{r_{2}})g(\mathbf{r_{2}},%
\mathbf{r_{3}})g^{\star }(\mathbf{r_{3}},\mathbf{r_{4}})g(\mathbf{r_{4}},%
\mathbf{r_{1}})=e^{\eta _{g}}$, where:%
\begin{equation*}
\eta _{g} = \frac{i}{2}\left( \left[ \mathbf{r_{1}}\times \mathbf{r_{2}}%
\right] -\left[ \mathbf{r_{2}}\times \mathbf{r_{3}}\right] +\left[ \mathbf{%
r_{3}}\times \mathbf{r_{4}}\right] -\left[ \mathbf{r_{4}}\times \mathbf{r_{1}%
}\right] \right) = 
2i\left( Q_{x}R_{y}-Q_{y}R_{x}\right) ,
\end{equation*}%
which depends only on the vectors $\mathbf{R,Q}$. \ 

The product of the four Landau orbitals, labeled by $n_{1}=n+s+t$, $%
n_{2}=n+s,n_{3}=n$ , and $n_{4}=n+t$ , is given by the following expression:%
\begin{equation*}
\varphi _{0n_{1}}(\mathbf{r}_{1})\varphi _{0n2}^{\ast }(\mathbf{r}%
_{2})\varphi _{0n_{3}}(\mathbf{r}_{3})\varphi _{0n_{4}}^{\ast }(\mathbf{r}%
_{4}) 
= \exp \left[ \eta _{\Delta }^{\left( m\right) }+\eta _{\Delta }^{\left(
sq\right) }+\eta _{\Delta }^{\left( lin\right) }+\eta _{\Delta }^{\left(
0\right) }\right]
\end{equation*}%
with:%
\begin{eqnarray*}
\eta _{\Delta }^{\left( m\right) } &=&i\sum_{j}\varepsilon _{j}x_{j}y_{j}=i 
\left[ 2Q_{x}R_{y}+2Q_{y}R_{x}+\left( D_{x}P_{y}+P_{x}D_{y}\right) \right] \\
\eta _{\Delta }^{\left( sq\right) } &=&-\sum_{j}y_{j}^{2}=-\left(
Q_{y}^{2}+4R_{y}^{2}+\left( D_{y}^{2}+P_{y}^{2}\right) \right) \\
\eta _{\Delta }^{\left( lin\right) } &=&\sum_{j}\left( i\varepsilon
_{j}q_{n_{j}}x_{j}-q_{n_{j}}y_{j}\right)  \\
&=&i\sum_{j}\left( \varepsilon _{j}n_{j}\right) R_{x}+iq_{0}\frac{1}{2}\left(
Q_{x}N_{4}-2D_{x}t-2P_{x}s\right) 
-q_{0}\left( R_{y}N_{4}-D_{y}s-P_{y}t\right) \\
\eta _{\Delta }^{\left( 0\right) } &=&-\frac{1}{4}q_{0}^{2}%
\sum_{j}n_{j}^{2}=-\frac{1}{2}q_{0}^{2}\left[ 2n^{2}+2ns+2nt+st+s^{2}+t^{2}%
\right]
\end{eqnarray*}%
and: $N_{4}\equiv \sum n_{j}=4n+2s+2t$. \ 

The last factor, $\widetilde{K}_{4}(\left\{ \mathbf{r}\right\} \sim \exp
\left( \eta _{G}\right) $, arising from the transitional invariant parts of
the Green functions, is independent of the center of mass coordinates:%
\begin{equation}
\eta _{G}=-\sum \mu _{j}\rho _{j}^{2}=-\left[ 
\begin{array}{c}
\left( \mu _{1}+\mu _{2}+\mu _{3}+\mu _{4}\right) Q^{2}+2\left( \mu _{2}-\mu
_{4}\right) \left( DQ\right) + \\ 
2\left( \mu _{1}-\mu _{3}\right) \left( PQ\right) +\left( \mu _{2}+\mu
_{4}\right) D^{2}+\left( \mu _{1}+\mu _{3}\right) P^{2}%
\end{array}%
\right] .
\end{equation}

The simplest integration to carry out, over $R_{x}$ , yields a non-vanishing
result only if $\sum \left( \varepsilon _{j}n_{j}\right) =0$, justifying the
parametrization of $n_{j}$\ chosen above. In this case $\int
dR_{x}=L_{x}=a_{x}N_{x}$. Next, the $R_{y}$-integration,%
\begin{equation}
\int \exp \left[ 4iQ_{x}R_{y}-4R_{y}^{2}-q_{0}NR_{y}\right] dR_{y} \\
= \frac{1}{2}\sqrt{\pi }\exp \left[ \frac{1}{16}q_{0}^{2}N^{2}-\frac{1}{2}%
iq_{0}NQ_{x}-Q_{x}^{2}\right]
\end{equation}%
leads to a space independent correction, $\frac{1}{16}q_{0}^{2}N^{2}$, which
removes the $n$-dependence of $\eta _{\Delta }^{\left( 0\right) }$: $\eta
_{\Delta }^{\left( 0\right) }+\frac{1}{16}q_{0}^{2}N^{2}=-\frac{1}{4}%
q_{0}^{2}\left( s^{2}+t^{2}\right) $. As a result, summation over Landau
orbitals is trivially done, yielding the total number of orbitals $%
\sum_{n}1=N_{y}$.

Combining the $Q$-dependent terms the corresponding integral is:%
\begin{equation}
\int d^{2}Q\exp \left[ -\beta _{0}\left[ Q^{2}+2\beta _{24}\left(
DQ\right) +2\beta _{13}\left( PQ\right) \right] \right] \\
= \frac{\pi }{\beta _{0}}\exp \left[ \beta _{0}\left( \beta
_{24}^{2}D^{2}+\beta _{13}^{2}P^{2}+2\beta _{24}\beta _{13}\left( DP\right)
\right) \right]
\end{equation}%
where 
\begin{equation}
\beta _{0}=\mu _{1}+\mu _{2}+\mu _{3}+\mu _{4}+1,\beta _{24}=\frac{\mu
_{2}-\mu _{4}}{\beta _{0}},\beta _{13}=\frac{\left( \mu _{1}-\mu _{3}\right) 
}{\beta _{0}}  \label{betas}
\end{equation}

The most complicated analytical part of the calculation, the $DP$%
-integrations, is now done by introducing the 4D vectors: 
\begin{equation}
Z=\left\{ D_{x},D_{y},P_{x},P_{y}\right\} ,L=q_{0}\left\{ -it,s,-is,t\right\}
\label{Z and L}
\end{equation}%
and the $\tau $ dependent $4\times 4$ matrix: 
\begin{equation} 
U =  \left( 
\begin{array}{cccc}
\left( \mu _{2}+\mu _{4}\right) -\beta \beta _{24}^{2} & 0 & -\beta \beta
_{24}\beta _{13} & -i/2 \\ 
0 & \left( \mu _{2}+\mu _{4}\right) -\beta \beta _{24}^{2}+1 & -i/2 & -\beta
\beta _{24}\beta _{13} \\ 
-\beta \beta _{24}\beta _{13} & -i/2 & \left( \mu _{1}+\mu _{3}\right)
-\beta \beta _{13}^{2} & 0 \\ 
-i/2 & -\beta \beta _{24}\beta _{13} & 0 & \left( \mu _{1}+\mu _{3}\right)
-\beta \beta _{13}^{2}+1%
\end{array}%
\right)  \label{U-matrix}
\end{equation}%
and then performing the resulting Gaussian integrations to have:

\begin{equation}
\int d^{4}Z\exp \left[ -Z^{T}UZ+LZ\right] =\frac{\pi ^{2}}{\sqrt{\det U}}%
\exp \left[ L^{T}U^{-1}L\right] .
\end{equation}
\end{widetext}

\end{document}